\documentclass[10pt,letterpaper]{article}

\usepackage[utf8]{inputenc}

\usepackage{graphicx}                       %
\usepackage{geometry}                       %
\usepackage{amsmath,amsfonts,bm}            %
\usepackage{enumitem}                       %
\usepackage{natbib}                         %
\usepackage[hidelinks, colorlinks=false, 
linkcolor=black, citecolor=blue]{hyperref}  %
\usepackage{fancyhdr, calc, lastpage}       %
\usepackage{booktabs}                       %
\usepackage{lipsum}                         %
\usepackage{endnotes}                       %
\usepackage{subfig}
\renewcommand\title[1]{{\linespread{1} \noindent\LARGE \bf \hskip2.25pc \parbox{.8\textwidth}{%
\LARGE \bf \begin{center} #1 \end{center}\rm } \rm\normalfont\normalsize} }
\renewcommand\author[1]{{\linespread{1} \noindent\hskip2.25pc \parbox{.8\textwidth}{%
   \normalsize \bf \begin{center} #1 \end{center}\rm } \vskip-1.4pc }}
\newcommand\address[1]{{\linespread{1} \noindent\hskip2.25pc \parbox{.8\textwidth}{%
   \footnotesize \it \begin{center} #1 \end{center}\rm }  \normalsize \vskip-1pc }}
\newcommand\email[1]{\vskip-.3cm \noindent\parskip0pc\hskip2.25pc \footnotesize%
   \parbox{.8\textwidth}{\begin{center}\it #1 \rm \end{center} } \normalsize  \vskip-.2cm}
\newcommand\PACS[1]{\vskip-2.75pc \begin{center}\parbox{.8\textwidth}{\small \copyright 2015 Acoustical Society of America\hfill \\ \small\bf PACS numbers: \rm #1 \hfill} \end{center}\vskip4pt}%

\renewenvironment{abstract}%
{\vskip1pc\noindent\begin{center} \begin{minipage}{.8\textwidth} {\bf Abstract: } }
{ \vspace{.25cm} \end{minipage}\end{center}\normalsize\vskip-1.5pc}%

\def\fps@table{h}%

\makeatletter
\newcommand\@MaxCapWidth{4.25in}
\long\def\@makecaption#1#2{%
  \small
  \vskip\abovecaptionskip
  \sbox\@tempboxa{#1. #2}%
  \ifdim \wd\@tempboxa >\@MaxCapWidth
    \hskip2.25pc\parbox{4.5in}{#1. #2}
  \else
    \global \@minipagefalse
    \hb@xt@\hsize{\hfil\box\@tempboxa\hfil}%
  \fi
  \vskip\belowcaptionskip\normalsize}
\makeatother

\makeatletter
\renewcommand\@seccntformat[1]{\csname the#1\endcsname.\hspace{.1cm}}

\renewcommand\section{\@startsection {section}{1}{0pt}%
                                     {-2ex plus -1ex minus -.2ex}%
                                     {0.65ex plus 1.2ex}%
                                     {\normalsize\bfseries}}
\renewcommand\subsection{\@startsection{subsection}{2}{0pt}%
                                     {-2.25ex plus -1ex minus -.2ex}%
                                     {.45ex plus .2ex}%
                                     {\normalsize\itshape}}
\renewcommand\subsubsection{\@startsection{subsubsection}{3}{0pt}%
                                     {-2.25ex plus -1ex minus -.2ex}%
                                     {1ex plus .2ex}%
                                     {\small\upshape}}
\makeatother

\makeatletter
\let\old@theendnotes\theendnotes
\renewcommand{\theendnotes}{\old@theendnotes\vspace{.3cm}}
\makeatother
\let\footnote=\endnote 
\begin{document}

\title{A model for the temporal evolution of the spatial coherence in decaying reverberant sound fields}

\author{Sam Nees, Andreas Schwarz, Walter Kellermann}
\address{Multimedia Communications and Signal Processing\\Friedrich-Alexander-Universität Erlangen-Nürnberg (FAU)\\Cauerstr. 7, 91058 Erlangen, Germany}
\email{samnees@gmail.com, schwarz@lnt.de, wk@lnt.de}
\begin{center}
\em Running title: Coherence in Decaying Reverberant Fields
\end{center}

\begin{abstract}
Reverberant sound fields are often modeled as isotropic. However, it has been observed that spatial properties change during the decay of the sound field energy, due to non-isotropic attenuation in non-ideal rooms. In this letter, a model for the spatial coherence between two sensors in a decaying reverberant sound field is developed for rectangular rooms. The modeled coherence function depends on room dimensions, surface reflectivity and orientation of the sensor pair, but is independent of the position of source and sensors in the room. The model includes the spherically isotropic (diffuse) and cylindrically isotropic sound field models as special cases.
\end{abstract}
\section{Introduction}

A reverberant sound field is generated by the reflections of an excitation signal at the boundaries of an enclosed environment. Common models for reverberant sound fields are isotropic sound field models which exhibit directionally invariant properties, with spherically isotropic (diffuse) or cylindrically isotropic sound fields being two well-known special cases \citep{cook_measurement_1955}. At any point within a spherically isotropic sound field, uncorrelated signals arrive from all directions in the three-dimensional space, whereas in a cylindrically isotropic sound field, signals arrive only from directions contained in a two-dimensional plane.
The use of an isotropic sound field to approximate a reverberant sound field can be justified as, after sufficient time, any signal originating from a single point in a reverberant environment will have many reflection paths to any other point within the sound field. These reflection paths will reach the point within the field from many different directions and will have high temporal densities. Additionally, if the source signal is also assumed to have limited temporal correlation, reflected signals arriving at the receiver can be assumed to be uncorrelated if the observation window length is limited \citep{jacobsen_coherence_2000}.

Isotropic sound field models are widely used to approximate reverberant sound fields in real rooms, e.g., for application to acoustic signal enhancement. However, rooms with strong variations between the acoustic properties of different surfaces, e.g., due to the use of curtains or acoustic ceiling tiles, can create a highly non-isotropic sound field with temporally varying spatial characteristics \citep{gover_measurements_2004}. 

In this letter, we investigate the temporal evolution of the spatial coherence function between two sensors in a decaying reverberant sound field in a rectangular room. Spatial coherence functions are an important measure for the properties of a sound field, and relevant for various signal processing applications, e.g., dereverberation by spatial postfiltering methods \citep{mccowan_microphone_2003,schwarz_coherent--diffuse_2015}. First, the well-known spherically isotropic (diffuse) and cylindrically isotropic models for idealized stationary sound fields are briefly reviewed. Then, we develop a specific model for the spatial coherence function of the short-time stationary decaying reverberant sound field in rectangular rooms. The spatial coherence function of the decaying sound field is found to be non-isotropic, time-variant, real-valued, and independent of the position of source and sensors within the room. The practical applicability of this model is finally evaluated by comparison with coherence estimates obtained from simulated and measured room impulse responses.

\section{Models for the spatial coherence of reverberant sound fields}

We consider a pair of omnidirectional microphones separated by a distance $d$. The sound field is assumed to be stationary. The spatial coherence function can be expressed as $\Gamma_{12}(k) = \frac{\Phi_{12}(k)}{\sqrt{\Phi_{11}(k) \Phi_{22}(k)}}$, where $k$ is the wavenumber, $\Phi_{11}(k)$, $\Phi_{22}(k)$ are the auto-power spectral densities (PSDs) of the two microphones signals, and $\Phi_{12}(k)$ is the cross-power spectral density (CPSD) between the two signals.

\begin{figure}[h!]
  \centering
  \includegraphics[width=0.5\textwidth]{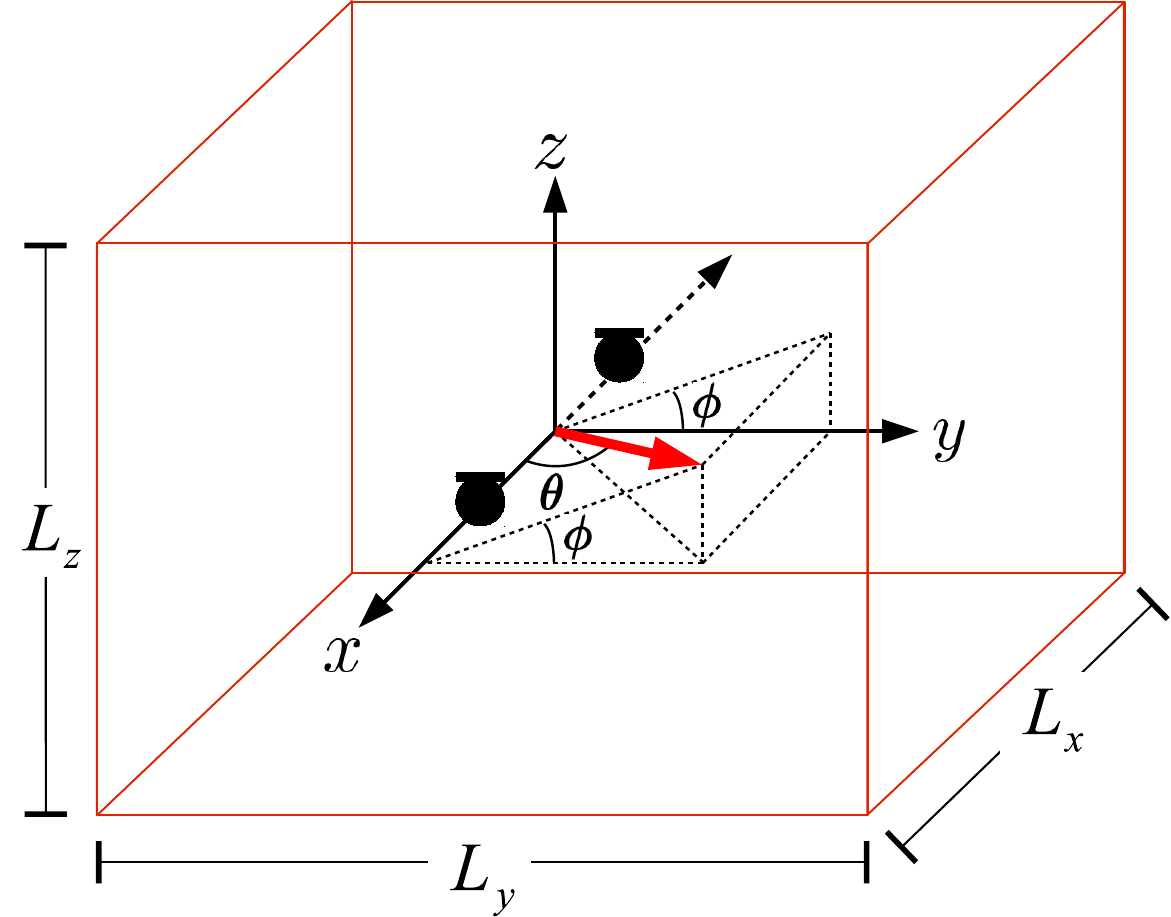}
  \caption{Spherical coordinates in rectangular environment}
  \label{fig:sphco}
\end{figure}

\subsection{Spherically isotropic (diffuse) sound field model}

In a spherically isotropic (diffuse) sound field, sound waves arrive uniformly from all directions in the 3-dimensional space. The PSD and CPSD can be calculated via surface integrals, where each point on the surface represents a direction from which a far-field signal is arriving at the microphone pair \citep{cook_measurement_1955}. With the spherical coordinates defined as shown in Fig.~\ref{fig:sphco}, and the microphone pair aligned with the $x$ axis, a signal arriving at the microphone pair from a given direction ($\theta$, $\phi$) will contribute to the CPSD with the phase term $e^{-jkd\cos(\theta)}$. To obtain the CPSD of a spherically isotropic sound field, this contribution can be inserted into a surface integral over a sphere. A surface integral over the PSD contributions yields the PSD, which is equal for both microphones ($\Phi_{11}(k)$ = $\Phi_{22}(k)$) since no obstructions are present and both microphones are assumed to be omnidirectional, and the spatial coherence function is obtained as \citep{cook_measurement_1955}:
\begin{align}
	\Gamma_{1 2}(k)
	&=	
	\frac{
	  \int_{0}^{2\pi} \int_{0}^{\pi} 
	  e^{-jkd \cos(\theta)}
	  \sin(\theta) 
	  \mathrm{d}\theta \mathrm{d}\phi  
	}
	{
	  \int_{0}^{2\pi} \int_{0}^{\pi} 
	  \sin(\theta) 
	  \mathrm{d}\theta \mathrm{d}\phi  
	}
	=
	\mathrm{sinc}(kd).
\end{align}

\subsection{Cylindrically isotropic sound field model}
For the case of a cylindrically isotropic sound field the coherence can be calculated similarly to that of a spherically isotropic field. However, it is now assumed that the sound waves only travel in the horizontal (x-y) plane, and the integration is therefore performed over the circumference of a circle in the x-y-plane, resulting in the zeroth-order Bessel function of the first kind \citep{cook_measurement_1955}:
\begin{align}
	\Gamma_{1 2}(k)
	&=	
	\frac{
	  \int_{0}^{2\pi}
	  e^{-jkd \cos(\theta)}
	  \mathrm{d}\theta
	}
	{
	  \int_{0}^{2\pi}
	  \mathrm{d}\theta
	}
	=
	J_{0}(kd).
\end{align}
This model was found to be a good approximation for reverberant sound fields in environments with strong attenuation along an axis perpendicular to the axis of the microphone pair, e.g., rooms with strongly absorbing floor and ceiling \citep{elko_superdirectional_2000,schwarz_coherent--diffuse_2015}.

\subsection{Proposed model for rectangular rooms}
While the spherically and cylindrically isotropic sound field models are often used to approximate the spatial coherence function of reverberant sound fields, real rooms rarely have isotropic properties. In practice, different reflection coefficients of surfaces create a non-isotropic reverberant sound field, with spatial properties which change during the decay of the sound field energy. We now develop a model for the time-varying spatial coherence of a sound field which is decaying from an initial impulse-like excitation at $t=0$. Unlike for the diffuse and cylindrically isotropic sound field models, which assume stationarity, the sound field is now only assumed to be short-time stationary, i.e., stationary within a small interval around the time $t$, with a spatial coherence $\Gamma_{12}(k, t)$ which is dependent on the time $t$ after the initial excitation.

In order to model sound field behavior within a rectangular room characterized by its dimensions and surface reflection coefficients, the ray-based sound model from geometrical room acoustics is employed. This model treats a sound field as being composed of incoherent rays which propagate only in straight lines and undergo no diffusion or interference during propagation and reflection \citep{kuttruff_room_2000}. Note that, although we neglect interference, we do consider the phase relations of a ray between the two microphones. Rectangular geometries along with the ray model have been previously applied to study the energy decay of sound fields, e.g., \cite{sakuma_approximate_2012, hirata_geometrical_1979}. However, to the authors' best knowledge, models for the spatial coherence of decaying sound fields in rectangular rooms have not been published.

The initial state of the sound field is modeled as spherically isotropic. This can be justified as, while the distribution of rays in a sound field is initially dependent on the source location, it quickly becomes spatially dispersed via reflections, obscuring that it originated from a single point. The time until this state can be assumed is known as the \textit{mixing time} and is typically assumed to be reached once 10 reflected signals are incident within a 24\,ms interval \citep{polack_acoustic_2008,jot_analysis_1997}. While it is dependent on room size, typical numbers for the mixing time are on the order of several 10\,ms.

While the sound field initially trends towards isotropy, it has been observed that isotropy is lost in the long term, since components along directions with stronger attenuation decay more rapidly \citep{gover_measurements_2004}. This effect will be covered by the model proposed in the following. %

A single sound ray arriving at a microphone pair has a velocity $c$, equal to the speed of sound, and a direction of propagation defined in the spherical coordinate system, shown in Fig.~\ref{fig:sphco}, by $\theta$ and $\phi$.  This velocity can be decomposed into its $x$, $y$ and $z$ components as follows:
 	\begin{align}
		c_x = c\cos(\theta), 
		c_y = c\sin(\theta)\cos(\phi),
		c_z = c\sin(\theta)\sin(\phi).
	\end{align}
The rectangular room in Fig.~\ref{fig:sphco} has dimensions defined by $L_x, L_y, L_z$. For a sound ray traveling in this room, the number of collisions with the two walls incident to the $x$ axis, $n_{x}$, which will have occurred after time $t$, is solely dependent on the absolute value of the component of its velocity in the $x$ direction, $|c_x|$, and can be formulated as:
	\begin{align}
		n_{x} = \left \lfloor{\frac{t |c_x|}{L_x}}\right \rfloor \simeq \frac{t |c_x|}{L_x}.
	\end{align}
Dropping the floor operator $\lfloor \cdot \rfloor$ is an approximation due to the fractional component which is included in addition to the correct integer number of collisions. However, as time $t$ increases, this fractional component represents a smaller and smaller percentage of the total number of collisions and can therefore be neglected.
Analogously, the number of collisions of a sound ray with the walls incident to the $y$ and $z$ axes in a given time period $t$ is given by $n_{y} \simeq \frac{t |c_y|}{L_y}, n_{z} \simeq \frac{t |c_z|}{L_z}$, respectively.
	
Now a reflection coefficient for each pair of parallel walls ($R_x, R_y, R_z$) is introduced, such that the initial power $A_0$ of a sound ray is reduced to $A_0R_x^2$ after a collision with one of the walls incident with the $x$ axis. After $n_x$ collisions with walls incident to the $x$ axis it can be seen that the power will be reduced to $A_0R_x^{2n_x}$.  After a given time period $t$ the total power reduction due to collisions can then be closely approximated by:
	\begin{align}
		A(\theta, \phi, t)
		= 
		A_0R_x^{\frac{2t |c_x|}{L_x}}R_y^{\frac{2t |c_y|}{L_y}}R_z^{\frac{2t |c_z|}{L_z}}
		=
		A_0
		R_x^{\frac{2t c|\cos(\theta)|}{L_x}}
		R_y^{\frac{2t c\sin(\theta)|\cos(\phi)|}{L_y}}
		R_z^{\frac{2t c\sin(\theta)|\sin(\phi)|}{L_z}}.
		\label{eq:ray_model}
	\end{align}
If it is desired to model the two walls incident to the $x$ axis with differing reflection coefficients, this can be approximated by setting $R_x$ equal to the geometric mean of the two. The resulting model error tends to zero as the number of reflections increases.

Eq.~\ref{eq:ray_model} provides a model for the PSD contribution of a sound ray, with initial PSD $A_0$, traveling in the direction $(\theta$, $\phi)$. Due to the initial diffuseness of the sound field, $A_0$ is assumed to be the same for rays propagating in all directions.
Now, in order to consider the contribution of this ray to the CPSD measured by a sensor pair with spacing $d$ and axis orientation $(\theta_\mathrm{mic}$, $\phi_\mathrm{mic})$, we define the phase term
\begin{align}
\rho(\theta, \phi, k) = \mathrm{e}^{-jkd(\cos(\theta_\mathrm{mic})\cos(\theta) + \sin(\theta_\mathrm{mic})\sin(\theta)\cos(\phi_\mathrm{mic} - \phi))}
\end{align}
representing the frequency-dependent phase difference between the sensor signals for a plane wave incident from the direction $(\theta, \phi)$. Assuming lossless propagation, the CPSD of this ray will then be equal to the PSD of the ray rotated by this phase term.

The CPSD and PSD of the sensor signals in the reverberant sound field can now be computed by surface integrals over a sphere to combine the contributions of rays incident from all directions. The spatial coherence of the decaying reverberant sound field at time $t$ is obtained as:
	\begin{align}
		\Gamma_{1 2}(k,t)
		&=	
		\frac{
			\Phi_{1 2}(k,t)
		}
		{
			\sqrt{\Phi_{11}(k,t) \Phi_{22}(k,t)}
		}
		=	
		\frac{
			\int_{0}^{2\pi} \int_{0}^{\pi} 
			\rho(\theta, \phi)
			A(\theta,\phi,t)
			\sin(\theta) 
			\mathrm{d}\theta \mathrm{d}\phi
		}
		{
			\int_{0}^{2\pi} \int_{0}^{\pi} 
			A(\theta,\phi,t)
			\sin(\theta) 
			\mathrm{d}\theta \mathrm{d}\phi
		}
		\\
		&=	
		\frac{
			\int_{0}^{2\pi} \int_{0}^{\pi} 
			\rho(\theta, \phi)
			R_x^{\frac{2t c|\cos(\theta)|}{L_x}}
			R_y^{\frac{2t c\sin(\theta)|\cos(\phi)|}{L_y}}
			R_z^{\frac{2t c\sin(\theta)|\sin(\phi)|}{L_z}}
			\sin(\theta) 
			\mathrm{d}\theta \mathrm{d}\phi
		}
		{
			\int_{0}^{2\pi} \int_{0}^{\pi} 
			R_x^{\frac{2t c|\cos(\theta)|}{L_x}}
			R_y^{\frac{2t c\sin(\theta)|\cos(\phi)|}{L_y}}
			R_z^{\frac{2t c\sin(\theta)|\sin(\phi)|}{L_z}}
			\sin(\theta) 
			\mathrm{d}\theta \mathrm{d}\phi
		}.
		\label{eq:coherence_model}
	\end{align}
Closed-form solutions of these integrals can be obtained for special cases. Most notably the coherence of a spherically isotropic field is obtained for $R_x = R_y = R_z =1$ and arbitrary $\theta_\mathrm{mic}$, $\phi_\mathrm{mic}$, and the coherence of a cylindrically isotropic field is obtained for $R_x = R_y = 1$, $R_z =0$, $\phi_\mathrm{mic} = 0^{\circ}$ and arbitrary $\theta_\mathrm{mic}$. The modeled coherence is independent of the source and microphone position, but it is dependent on the orientation of the microphone pair in the room. Microphone pairs aligned with the $x$ axis ($\theta_\mathrm{mic} = 0^{\circ}$), as shown in Fig.~\ref{fig:sphco}, result in the simplified phase term $\rho(\theta, \phi, k) = \mathrm{e}^{-jkd\cos(\theta)}$.

Note that the coherence generated from this model will always be real-valued, independently of the orientation of the sensor pair. This is due to the fact that walls are parallel, and therefore the imaginary part of the phase term of a ray arriving from the direction $(\theta, \phi)$ is always exactly cancelled by that of a symmetric ray from the opposite direction $(\pi-\theta, \phi+\pi)$, i.e., $\Im\{\rho(\theta, \phi, k)\}+\Im\{\rho(\pi-\theta, \pi+\phi, k)\}=0$, where $\Im\{\cdot\}$ denotes the imaginary part of a complex number.
\section{Evaluation}

First, we compare the proposed model to coherence estimates from impulse responses generated by the image source method \citep{allen_image_1979,peterson_simulating_1986,habets_room_2010}. An environment of dimensions $[L_x, L_y, L_z] = [6, 4, 3]\mathrm{m}$ was simulated containing a linear array of 16 microphones, with uniform spacing, $d=0.08\,\mathrm{m}$, aligned with the $x$ axis and centered in the room. The source was located at $[4.8, 3.2, 2.6]\,\mathrm{m}$ with the origin set at the bottom left hand back corner of the room. The reflection coefficients of the walls incident to the $x$ axis were set to $R_x=0.8$; for the other walls the coefficients were set to $R_y=R_z=1.0$. An impulse response was generated from a source position to each microphone, with a sampling rate of 16\,kHz, and split into non-overlapping intervals 100\,ms in length, in order to allow evaluation of the time-variant characteristics of the decaying sound field. In each interval the signal was short-time Fourier transformed using a window length of 1024, 75\% overlap and a DFT of size 512. A single spatial coherence estimate for each time interval was then obtained by averaging the involved spectra over the entire interval and over all adjacent microphone pairs. This was done to reduce the variance of the coherence estimate, allowing the accuracy of the model to be more easily evaluated. The proposed coherence function (Eq.~\ref{eq:coherence_model}) was then computed using numerical integration in MATLAB. For each interval, the time point $t$ for the computation of the coherence model was chosen 10\,ms from the start of the interval, since the average coherence over the interval is dominated by the earlier contributions, when the field has more energy and therefore provides a larger contribution to the averaged coherence estimate.

\begin{figure}[h!]
  \centering
  \includegraphics[width=0.9\textwidth]{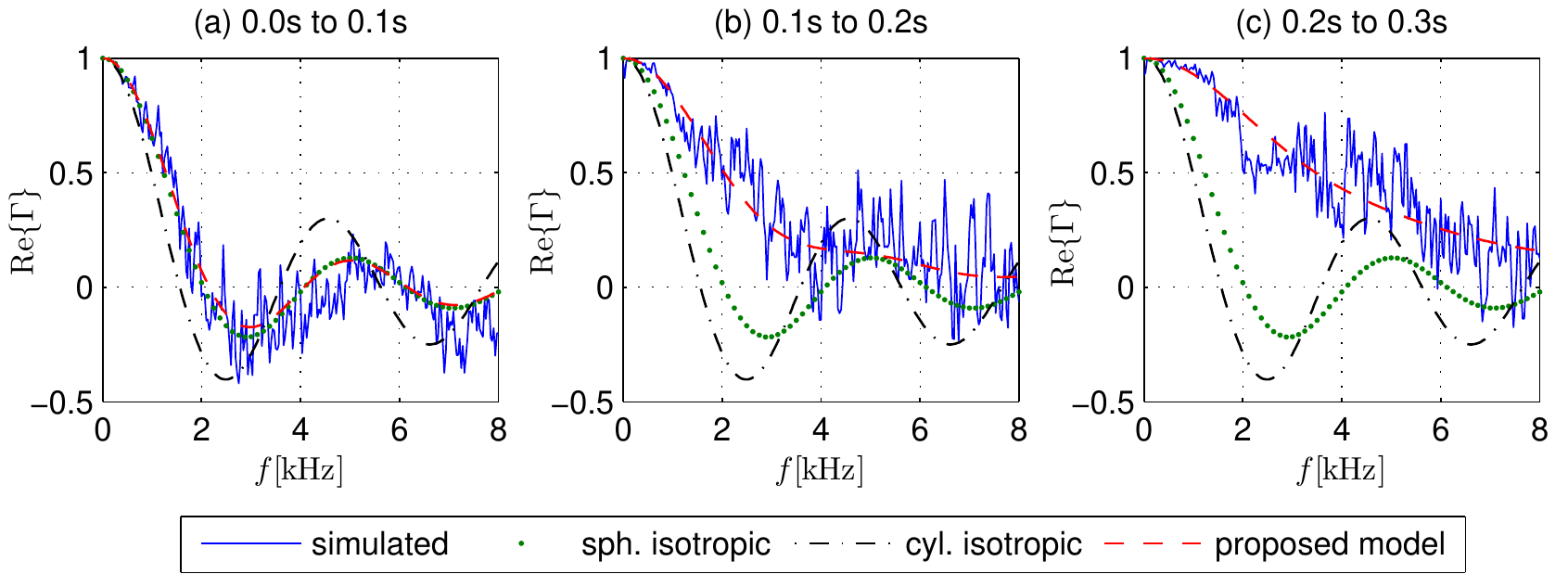}
	\caption{Comparison of coherence simulated using the image source model and coherence calculated using the presented model (Eq.~\ref{eq:coherence_model}) for sensor orientation $\theta_\mathrm{mic} = 0, \phi_\mathrm{mic} = 0$, within time intervals (a) 0.0\,s to 0.1\,s, (b) 0.1\,s to 0.2\,s, (c) 0.2\,s to 0.3\,s.}
	\label{fig:cohcompal}
\end{figure}

\begin{figure}[h!]
  \centering
  \includegraphics[width=0.9\textwidth]{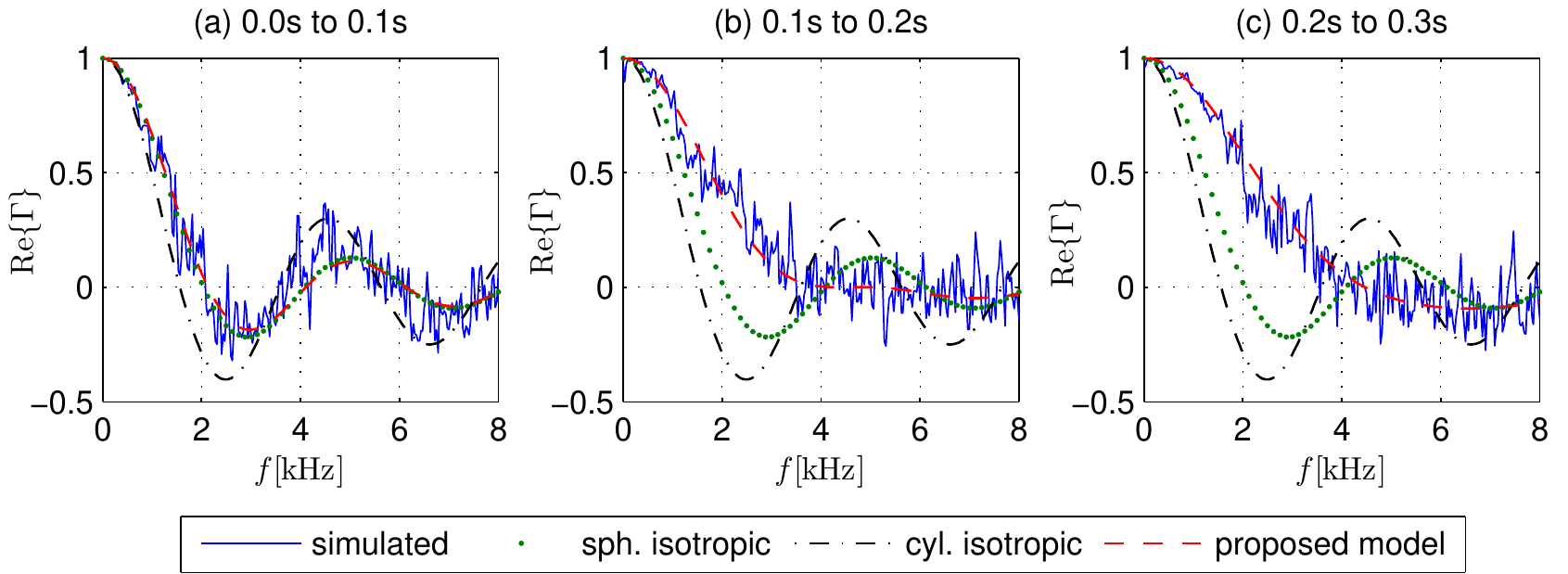}
	\caption{Comparison of coherence simulated using the image source model and coherence calculated using the presented model (Eq.~\ref{eq:coherence_model}) for sensor orientation $\theta_\mathrm{mic} = 20^{\circ}, \phi_\mathrm{mic} = 36^{\circ}$, within time intervals (a) 0.0\,s to 0.1\,s, (b) 0.1\,s to 0.2\,s, (c) 0.2\,s to 0.3\,s.}
	\label{fig:cohcompunal}
\end{figure}

Fig.~\ref{fig:cohcompal} and Fig.~\ref{fig:cohcompunal} show the coherence of the simulated room impulse response for two different orientations of the microphone array, in comparison to the ideal isotropic models and the proposed model. Note that the fluctuations in the simulated coherence are due to the estimation variance. While the sound field is isotropic for the first interval, it becomes increasingly anisotropic as the sound field decays; this behavior is matched by the proposed model. Note that the model is also able to account for the effect of sensor orientation. A similar agreement between the presented model and simulated impulse responses was also obtained for other configurations of the microphone array and room parameters.

\begin{figure}[h!]
  \centering
  \includegraphics[width=0.9\textwidth]{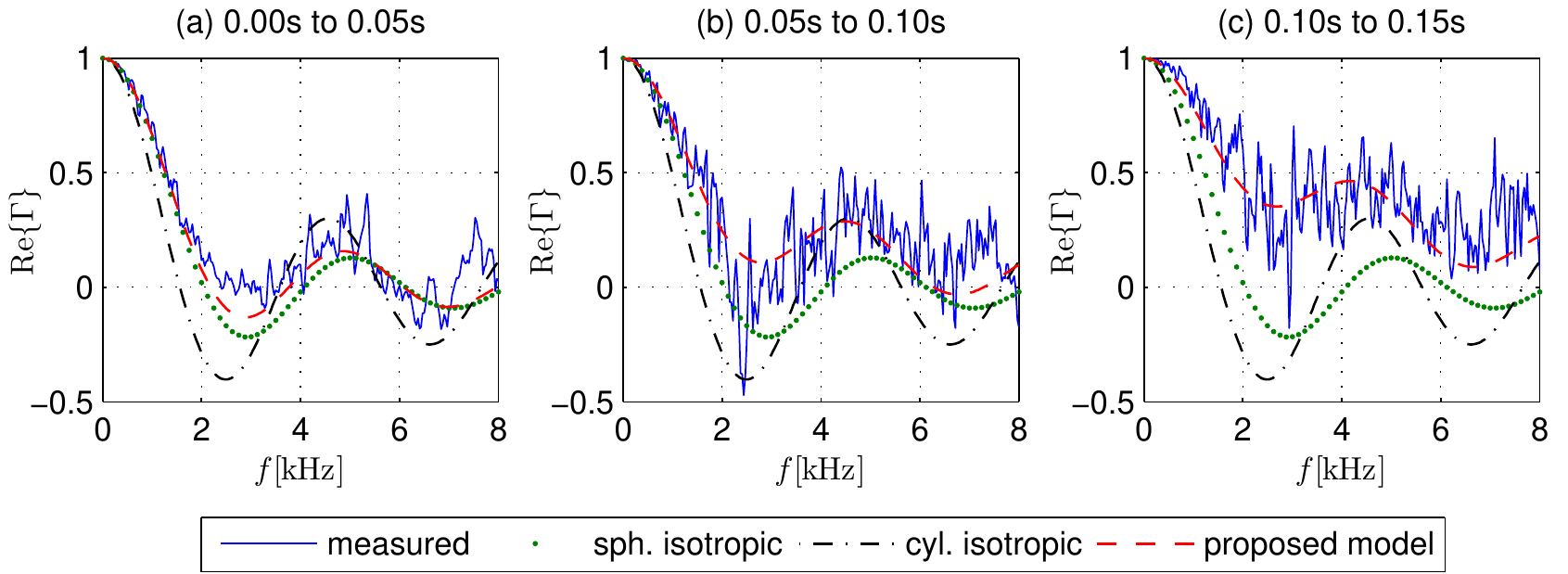}
	\caption{Comparison of coherence from measured room impulse responses and coherence calculated using the presented model (Eq.~\ref{eq:coherence_model}) within time intervals (a) 0.0\,s to 0.05\,s, (b) 0.05\,s to 0.1\,s, (c) 0.1\,s to 0.15\,s.}
	\label{fig:measuredcohcomp}
\end{figure}

The coherence model was also compared to measured data in physical environments. This was done by measuring impulse responses in a room, from which the coherence was then estimated within non-overlapping 0.05\,s time intervals, using the same techniques as for the simulated impulse responses, however only averaging over two microphone pairs. The room dimensions were 6m x 6m x 3m, with a reverberation time $T_{60} \approx 0.4\,\mathrm{s}$, and the reflection coefficients of the room surfaces were estimated to be $R_x = 0.60$, $R_y = 0.65$, $R_z = 0.80$, using typical material properties. The presented coherence model was then used with these parameters to model the coherence of the decaying sound field in the room, and compared to the coherence estimated from the impulse responses. Fig.~\ref{fig:measuredcohcomp} shows that the measured coherence initially roughly matches that of the spherically isotropic model as expected, but then diverges from that model. The proposed model matches the measured coherence well for all time intervals.

\section{Conclusions}

A model for the temporal evolution of the spatial coherence between two sensors in a decaying reverberant sound field was developed, based on a ray-based sound propagation model in a rectangular room. Using room dimensions, surface reflection coefficients and orientation of the sensor pair as known parameters, the model can predict the time-varying characteristics of the coherence function, as confirmed by experiments using simulated and measured room impulse responses. The modeled coherence function is independent of the source and sensor positions in the room, and, due to the rectangular room geometry, always real-valued. For special cases, the proposed model simplifies to the ideal spherically and cylindrically isotropic sound field models.

The presented model is based on a reverberant sound field generated by reflection and attenuation in a rectangular room, where each surface is characterized by a single reflection coefficient. More complex geometries, non-uniform surface reflectivity factors, or diffracting objects obstructing the sound field may impact the applicability of this model.

\bibliography{jasa-el_cleaned}
\bibliographystyle{jasa}

\end{document}